# Advanced Distributed Submarine Cable Monitoring and Environmental Sensing using Constant Power Probe Signals and Coherent Detection

Mikael Mazur (1), Nicolas K. Fontaine (1), Megan Kelleher (2,3), Valey Kamalov (4), Roland Ryf (1), Lauren Dallachiesa (1), Haoshuo Chen (1), David T. Neilson (1) and Franklyn Quinlan (2,3)

(1) Nokia Bell Labs, 600 Mountain Ave, New Providence, NJ 07974, USA
(2) Department of Physics, University of Colorado Boulder, Boulder, Colorado 80309, USA
(3) National Institute of Standards and Technology, Boulder, Colorado 80305, USA
(4) Optica, 2010 Massachusetts Ave. NW Washington, DC 20036 USA
Email: mikael.mazur@nokia-bell-labs.com

**Abstract:** In this work we demonstrate an FPGA-based coherent optical frequency domain reflectometry setup for cable monitoring. Using coherent detection for averaging and narrowband filtering, we significantly improve the signal-to-noise ratio (SNR) compared to traditional intensity-only techniques while also enabling continuous monitoring of phase and polarization. In addition, the probe signal has constant power, avoiding the nonlinear distortions and thereby enabling continuous use without interfering with data channels. We perform a field demonstration over a trans-oceanic cable using loopback configurations present in each repeater, demonstrating measurement SNR exceeding 30 dB for all about 80 repeaters with an averaging window of 1 second. We furthermore compare cable monitoring using coherent processing to today's solutions using power-only measurements. Our results show how transitioning to coherent technology for cable monitoring can improve sensitivity and enable new types of monitoring, exploring knowledge gained from transitioning from incoherent to coherent data transmission.

## 1. Introduction

Submarine cables form the backbone of the Internet, enabling high-speed connectivity between countries and continents [1]. Traffic interruptions caused by cable breaks are costly and cable repair is a slow tedious process[2]. Given the rapid growth in both the number of cables, and the capacity per cable, active monitoring to detect early signs of a cable break is vital to improve the robustness of future systems [3]. Today, cable monitoring is typically performed using passive monitoring of repeaters via time-domain pulsed signals. Cable break localization is performed in a similar fashion using high-power pulses. However, for continuous in-service monitoring during operation, the pulse energy must be minimized to avoid non-linear distortions. This leads to long averaging times, which is acceptable for amplifier monitoring, but prevents continuous cable monitoring and environmental sensing. More recently, the potential for exploiting submarine cables as environmental sensing tools have been proposed. Monitoring of ocean currents, earthquakes, swells have been demonstrated using both distributed acoustic sensing (DAS)[4,5], state-of-polarization (SOP)[6-8] and laser phase interferometry [8-9]. DAS enables distributed sensing using Rayleigh backscattering but is typically limited to a reach of about 170km [10]. While these systems enable very high sensitivty and spatial localization, this limits their use to monitoring close to the cable endpoints. SOP and laser interferometry can be done using either dedicated probe signals [9] or using coherent transceivers [6,8]. In the latter case, the digital signal processing (DSP) used to compensate channel distortions is effectively turned into a sensor by monitoring the tracking algorithms. Distributed polarization and phase measurements have been shown using high loss loop-back (HLLBs) present in submarine cables for cable monitoring purposes. The SOP demo used time-domain pulses generated from a coherent transceiver and a stokes receiver[11]. Similarly to the cable monitoring system, this approach suffers from lower SNR dictated by the limited pulse energy acceptable and long dead-times between measurements given by the cable roundtrip propagation time. Marra et. al., demonstrated phase-sensing using fiber Bragg gratings (FBGs) and frequency sweeps, enabling the use of continuous probing signals [12]. However, the use of analog down conversion only allowed monitoring of a single repeater at the time and the system only used single polarization signals.

In this work we present a real-time FPGA-based long-range optical frequency domain reflectometry (OFDR) [13] setup for simultaneous cable monitoring and fiber sensing. We use a dual-polarization constant-power probe signal and coherent detection combined with real-time DSP processing to enable zero-deadtime continuous monitoring of both phase and polarization changes for all repeaters along the cable in parallel. We show how the proposed system can be used for simultaneous cable monitoring and fiber sensing, demonstrated via monitoring. We focus our analysis of the deep-ocean region beyond the first span of the cable, targeting a regime beyond what is widely monitored using commercially available traditional DAS systems. Our results highlights the potential for future cable monitoring systems to improve cable integrity and enable new applications of submarine cables.

## 2. Experimental Setup

The experimental setup is shown in Fig. 1. The advanced cable monitoring prototype, shown in Fig. 1b consists of three main building blocks: A laser source, an FPGA-based coherent signal modulation and detection stage and a GPU-based real-time processing unit. The laser source consisted of a fiber laser with a nominal linewidth below 100Hz. The fiber laser had active stabilization via locking to a filter inside the laser package. For performance comparison we also compared this to an external cavity stabilized laser system [14]. The cavity was under vacuum (approx 1e-8 Torr), placed on a vibration isolation table on the floor and surrounded by multiple layers of acoustic foam. The main laser was placed in a rack box and screwed into a regular telecom rack. The continuous power dual-polarization probe signal was generated using a regular dual-polarization IQ-modulator. The driving signals were digitally generated using an FPGA (Xilinx ZU49DR) with 14-bit digital-to-analog converters operating at a clock frequency of 125MHz to generate four 6GS/s signals. The electrical driving signals were phase-locked to enable stable single-sideband modulation. The sweep bandwidth was 125MHz, centered at an intermediate frequency of 500MHz. The modulated output was amplified and filtered before fed into the cable monitoring input port. The receiver consisted of polarization-diverse heterodyne receiver using the transmitter laser as local oscillator. The two output signals were sampled at 2GS/s using 14-bit analog-to-digital converters. The digitized signals were stored in intermediate DDR RAM-based buffers on the FPGA and packaged into Remote Direct Memory Access version 2 (RDMAv2) Ethernet packets. These packets were continuously streamed over a regular 100Gbit/s Ethernet connection to the processing unit. The benefit on this approach is that the processing unit can be located elsewhere, or put on the cloud, making the system very flexible. One processing unit can also handle multiple interrogators and load distribution can easily be implemented based on regular Ethernet networking. The processing unit directly (without use of the CPU) placed the incoming data into memory accessible by the GPU. The receiver DSP consisted of digital demodulation of the incoming signals to extract the phase and full Jones matrix of each repeater along the cable. Compared to previous approaches using analog demodulation prior to detection, this enables continuous parallel monitoring of all repeaters along time cable. Intermediate positions can also be monitored using the Rayleigh backsacttering at the expense of increased data flow. A measurement bandwidth of 300kHz was used. The processed results were stored in files and uploaded to the cloud, requiring a bandwidth of only 80Mbit/s. This bandwidth can further be reduced by accounting for the target monitoring purpose such as slow monitoring to trigger tsunami warning alarms or more rapid events such as detecting anchor drag of a submarine cable.

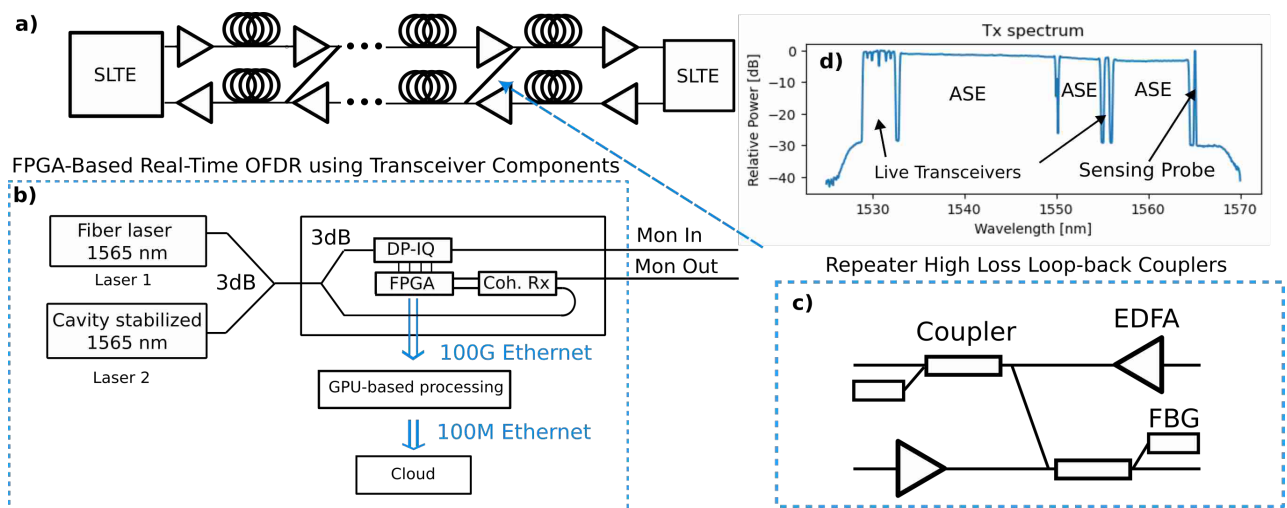

*Figure 1: (a) Schematic of a submarine transmission cable system using loop-back connections for monitoring. (b) Block diagram of the coherent continuous power dual-polarization sensing system. (c) Schematic of the high loss loop-back couplers present in each repeater. (d) Transmitted spectrum showing the probe signal out of band.*

The OFDR prototype system is connected to the input/output monitoring ports of the subsea cable. The fiber cable uses pairs of single-mode fiber for bi-directional transmission. It is transatlantic cable connecting North America with Europe. About 80 repeaters are required to amplify the signal and cable is equipped with high-loss loop-back configurations in each repeater, shown in Fig. 1c, allowing backscattered light from the forward transmission span to be coupled into the fiber going to the opposite direction. The transmitted spectrum is shown in Fig. 1d. The transmitted signal consisted of 7 live transceivers, amplified spontaneous emission (ASE) noise and the sensing probe placed at the monitoring wavelength outside the transmission band.

3. Results

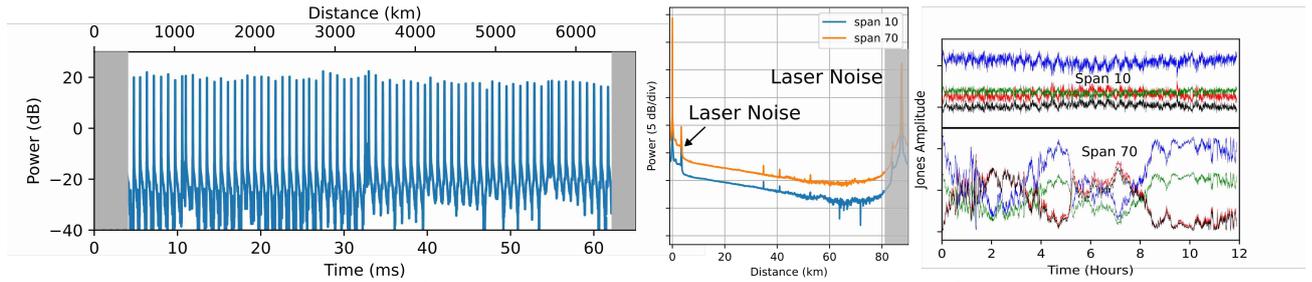

*Figure 2: (a) Measured trace from the submarine cable. (b) Zoom-in for intensity measurements from span 10 and span 70. The average time is 60 seconds. (c) All Jones components from the same spans.*

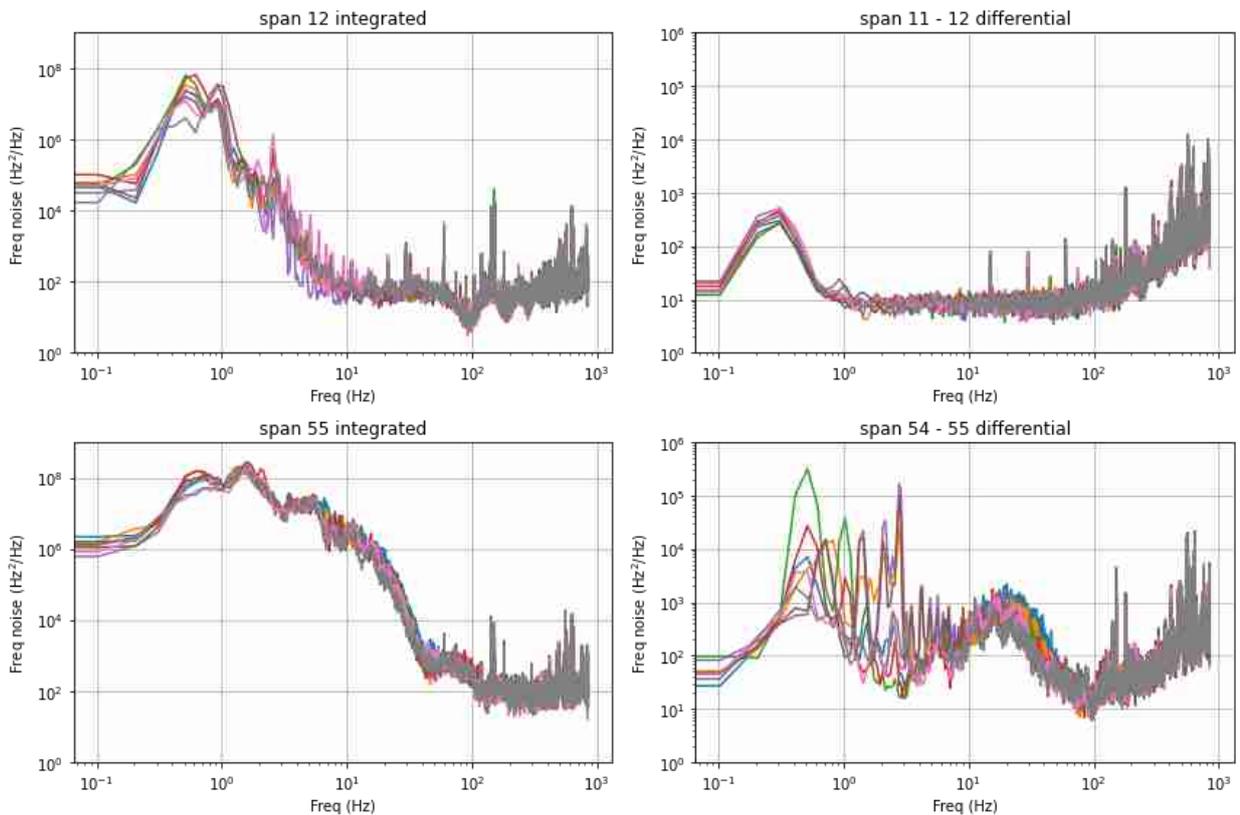

*Figure 3: Comparison of integrated and differential phase changes for span 12 and span 55 using the free-running fiber laser. Each measurement trace is 1 hour apart. We observe how the integrated amount of noise grows with distance, as expected, and covers a broad range of frequencies after 55 spans. Differential measurements enabling separation of effects, showing that span 12 is a very quiet span. In contrast, the span between repeaters 54 and 55 shows strong frequency components, matching well with previously observed Hz-level cable oscillations over transatlantic cables [8].*

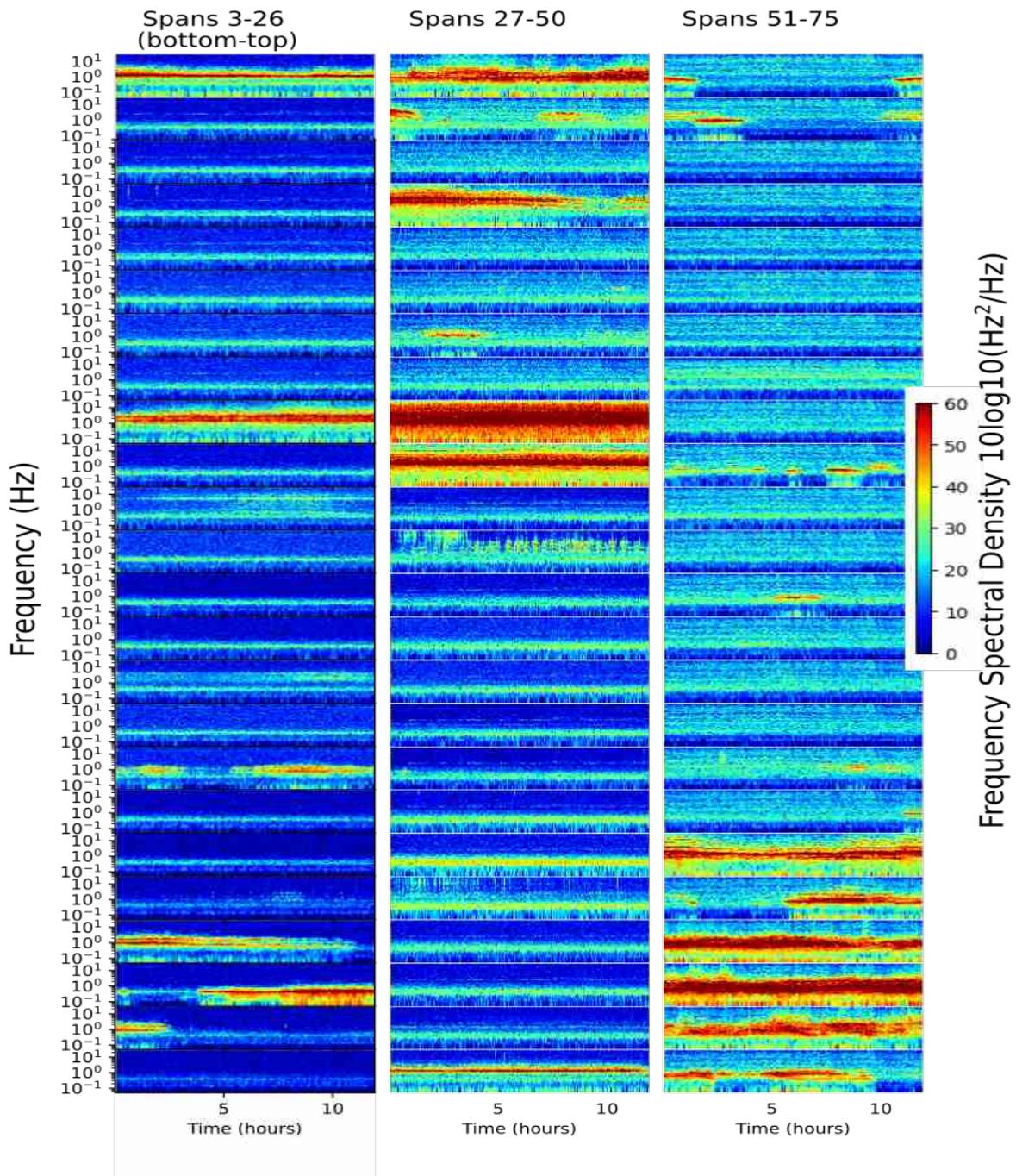

*Figure 4: Spectrograms showing the different spans over a 12h measurement time period covering 0.1 to 10Hz.*

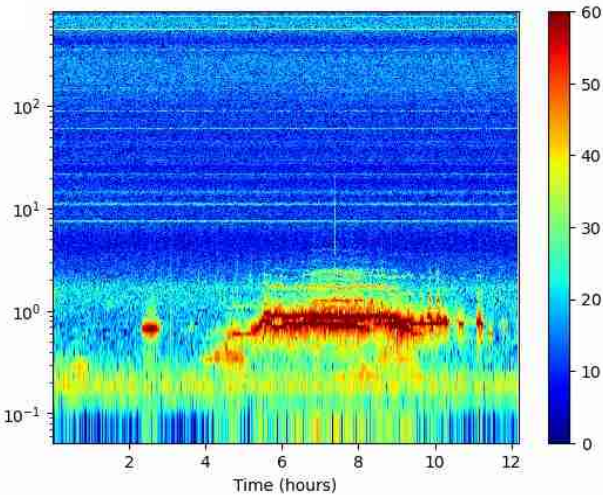

*Figure 5: Spectrogram of the phase from span 6. Oscillations observed from 0.1Hz up to a a few Hz observed around hour 5-10are in line with previous observations using polarization measurements from coherent transceivers.*

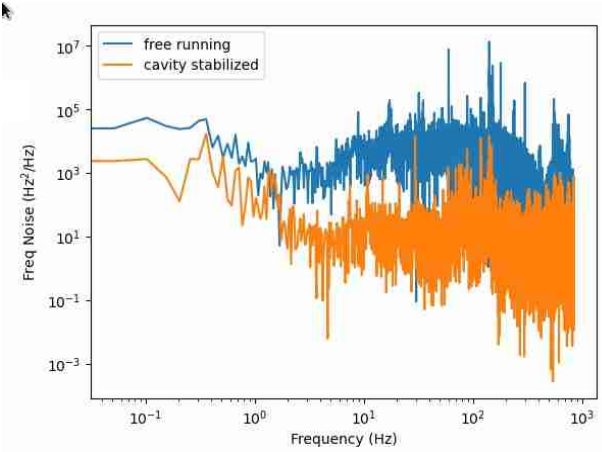

*Figure 6: Comparison of frequency noise between a free-running NKT fiber laser and the cavity-stablized laser system. The distance is about 400km. For frequencies below 1Hz, an improvement of around 10dB is observed, which increases to about 20dB for frequencies between 10Hz to 1kHz.*

Figure 2(a) shows a recovered time-domain trace from the OFDR system. Each repeater is clearly visible, with a resulting signal-to-noise (SNR) ratio of about 30dB over a measurement bandwidth of 300kHz. Intensity measurements averaged over 1 minute for two selected spans is shown in Fig. 2(b). This shows that intermediate features between repeaters can be accurately resolved at time-scales of minutes with high SNR. Using The corresponding four Jones elements making up the 2x2 Jones matrix are shown in Fig. 2(c).

The integrated and differential phase for span 12 and 55 are shown in Fig. 4. The traces are taken one hour apart. We clearly see the power of the distributed measurement technique. We observe the expected growth in broadband low frequency noise from span 12 to span 55. Looking at the differential noise, we see that span 12 is very quiet. In contract, span 55 shows a strong time dependence with clear spikes around Hz level. These oscillations match well with previously observed oscillations and are likely due to cable oscillations. A spectrogram of the phase from span 6 is shown in Fig. 5.

A frequency noise comparison between the free-running NKT fiber laser (X15) and the cavity-stabilized laser system is shown in Fig. 6. We observe about 10dB improvement for frequencies below 1Hz and up to 20dB improvement for frequencies in the 1Hz-1kHz regime. While the X15 laser is sufficient for differential measurements, and the main laser source used in this experiments, our results show that introducing cavity-stabilized systems can further improve the system sensitivity. This will likely be needed to reach to perform detection of weak low-frequency events such as deep-ocean tsunamis. Future long-reach OFDR systems is therefore likely gain significantly from developments of both portable reference cavities and ultra stable laser sources based on integrated photonics [15,16].

Using the proposed OFDR technique, the relative change in propagation time between different reflectors can be measured. An example of this is shown in Fig. 7, showing the relative movement between selected reflectors. Note that all reflectors are measured continuously in parallel but only a few are shown to ease visualization. A few interesting observations can be made from Fig. 7. First, the variation between different spans is quite large. Moreover, comparing for example span 10 and 11, we observe that the curves seem to move in opposite direction. This implies that reflector 10 is moving towards/away from reflector 11, respectively, which indicates a stretch/contraction of this fiber segment. While there might be different causes for this movement, such as changes in temperature or deep-ocean currents, it still demonstrates the powerful potential of continuous monitoring using OFDR. For example, span 20-22 are in general very quiet, even over days of measurements. These spans also map to a point of the geo-profile where the fiber is expected to lay flat on the sea bottom. More movement is observed for span 53-56, which roughly map to the mid-atlantic ridge. During this stretch, the fiber pass more rocky train and is expected to be more susceptible to changes.

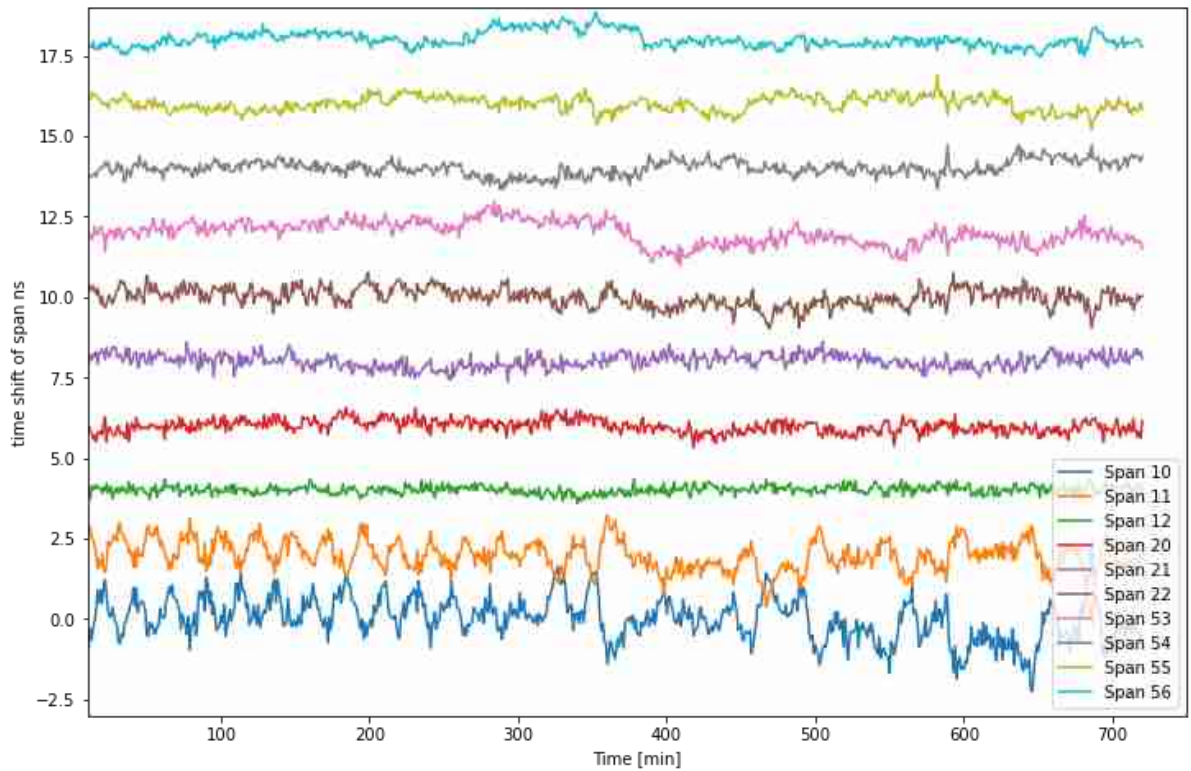

*Figure 7: Relative movement of different spans in ns extracted by monitoring the position of each grating. This monitor technique enables identification of cable movement. Looking at span 10-11, we observe highly correlated but opposite direction movement of the two repeaters.*

## 4. Conclusions

We have demonstrated a long-range OFDR fiber sensing built from regular telecommunication transceiver hardware, an FPGA for real-time signal generation and digitization and a GPU for post-processing. We performed a field trial over a transatlantic submarine fiber cable to demonstrate both phase and polarization sensing using continuous power probe signals and coherent detection. The use of constant-power probe signals avoid non-linear penalties and the use of coherent detection enables filtering and processing using digital signal processing to maximize the received signal quality. We demonstrated phase and polarization sensing covering sub-Hz to kHz range, enabling simultaneous sensing for both environmental monitoring and cable protection. The prototype is built from regular telecom transceiver components, showing the potential for transceiver hardware to act as the base for future dedicated sensing systems, reducing cost and easing scalability.

## 5. Acknowledgement

The authors would like to thank Rodney Dellinger and Benoit Kowalski from Nokia for experimental assistance.